\documentclass[%
 reprint,
 superscriptaddress,
 bibnotes,
 amsmath,amssymb,
 aps,
 prb,
 floatfix,
]{revtex4-2}

\usepackage{graphicx}
\usepackage{dcolumn}
\usepackage{bm}
\usepackage{hyperref}
\usepackage{float}
\usepackage[normalem]{ulem}
\usepackage{soul}
\usepackage{color}
\usepackage[T1]{fontenc}
\hypersetup{
  colorlinks   = true,
  urlcolor     = black,
  linkcolor    = black,
  citecolor   = blue
}
\usepackage{etoolbox}
\apptocmd{\sloppy}{\hbadness 10000\relax}{}{}
\usepackage{comment}

\usepackage{xcolor}

\begin{document}
\title{X-ray magnetic circular dichroism evidence of \\ intrinsic $d$-wave altermagnetism in rutile-structure NiF$_2$}


\author{Zezhong Li}
\thanks{Equal contributions}
\affiliation{National Synchrotron Radiation Laboratory and School of Nuclear Science and Technology, University of Science and Technology of China, Hefei 230026, China} 

\author{Kosuke Sakurai}
\thanks{Equal contributions}
\affiliation{Department of Physics and Electronics, Graduate School of Engineering, Osaka Metropolitan University, 1-1 Gakuen-cho, Nakaku, Sakai, Osaka 599-8531, Japan}

\author{Yiu-Fung Chiu}
\thanks{Equal contributions}
\affiliation{Diamond Light Source, Harwell Campus, Didcot, OX11 0DE, United Kingdom}
\affiliation{Department of Physics, University of Oxford, Clarendon Laboratory, Parks Road, Oxford OX1 3PU, United Kingdom}

\author{Dirk Backes}
\affiliation{Diamond Light Source, Harwell Campus, Didcot, OX11 0DE, United Kingdom}

\author{Dharmalingam Prabhakaran}
\affiliation{Department of Physics, University of Oxford, Clarendon Laboratory, Parks Road, Oxford OX1 3PU, United Kingdom}

\author{Mizuki Furo}
\affiliation{Department of Physics and Electronics, Graduate School of Engineering, Osaka Metropolitan University, 1-1 Gakuen-cho, Nakaku, Sakai, Osaka 599-8531, Japan}

\author{Choongjae Won}
\affiliation{Laboratory for Pohang Emergent Materials and Max Plank POSTECH Center for Complex
Phase Materials, Department of Physics, Pohang University of Science and Technology, Pohang
37673, Korea}

\author{Wenliang Zhang}
\affiliation{National Synchrotron Radiation Laboratory and School of Nuclear Science and Technology, University of Science and Technology of China, Hefei 230026, China} 

\author{Sang-Wook Cheong}
\affiliation{Keck Center for Quantum Magnetism, Department of Physics and Astronomy, Rutgers University, Piscataway, New Jersey, 08854, USA} 

\author{Andrew Boothroyd}
\affiliation{Department of Physics, University of Oxford, Clarendon Laboratory, Parks Road, Oxford OX1 3PU, United Kingdom}

\author{Mirian Garcia-Fernandez}
\affiliation{Diamond Light Source, Harwell Campus, Didcot, OX11 0DE, United Kingdom}

\author{Sahil Tippireddy}
\affiliation{Diamond Light Source, Harwell Campus, Didcot, OX11 0DE, United Kingdom}

\author{Jan Kune\v{s}}
\affiliation{Department of Condensed Matter Physics, Faculty of Science, Masaryk University, Kotl\'{a}\v{r}sk\'{a}  2, 61137 Brno, Czech Republic}

\author{Stefano Agrestini}
\email{stefano.agrestini@diamond.ac.uk}
\affiliation{Diamond Light Source, Harwell Campus, Didcot, OX11 0DE, United Kingdom}

\author{Atsushi Hariki}
\email{hariki@omu.ac.jp}
\affiliation{Department of Physics and Electronics, Graduate School of Engineering, Osaka Metropolitan University, 1-1 Gakuen-cho, Nakaku, Sakai, Osaka 599-8531, Japan}

\author{Ke-Jin Zhou}
\email{kjzhou@ustc.edu.cn}
\affiliation{National Synchrotron Radiation Laboratory and School of Nuclear Science and Technology, University of Science and Technology of China, Hefei 230026, China}

\date{\today}

\begin{abstract}
We present the x-ray magnetic circular dichroism (XMCD) at the Ni $L_{2,3}$-edge as an evidence of the $d$-wave altermagnetism in rutile-structure NiF$_2$. Sizable XMCD signal is observed in excellent agreement with theoretical simulations. Owing to a considerable net magnetization 
due to spin canting, the XMCD spectrum consists of an altermagnetic signal as well as a non-negligible ferromagnetic contribution. We verify experimentally that the XMCD spectrum can be written as a sum of contributions from altermagnetism and weak ferromagnetism.  
Two experimental methods to isolate the ferromagnetic contribution are shown to yield essentially the same result. 
These are dependence of XMCD on applied magnetic fields below the Néel temperature and
the XMCD measured in applied field above the Néel temperature. Our results demonstrate the utility of XMCD as a probe for altermagnetic materials with the coexisting weak ferromagnetism induced by the relativistic spin-orbit coupling.
\end{abstract}

\maketitle

Altermagnets have recently been classified as a distinct  group of collinear compensated magnets, opening a new frontier for exploring time reversal symmetry ($\mathcal{T}$) breaking responses, such as the anomalous Hall effect (AHE)~\cite{Naka2020,Smejkal2022C,Smejkal2020,Hayami2021,Mazin2021,Naka2022,cheong2025altermagnetism}, odd magneto-optical effects~\cite{Hariki2024,Sasabe2023,Zhou2021,Hariki2024b}, and spin polarized bands~\cite{Ahn2019,Hayami2019,Smejkal2020,Smejkal2020b,Yuan2020,Yuan2021,Hayami2020,Smejkal2020b,Mazin2021,Liu2022,Jian2023}, which were previously regarded as exclusive to ferromagnets. The crystal symmetries compatible with altermagnetism have been identified, yielding a growing number of candidate materials. However, the number of materials with experimentally confirmed altermagnetic signatures remains limited.

Rutile structure has been regarded as a prototypical host of altermagnetism as RuO$_2$ stood in the center of early studies thanks to theoretical reports of $d$-wave spin splitting~\cite{Ahn2019,Smejkal2020}.  Recent reports, however, have questioned the existence of magnetic order in RuO$_2$~\cite{Hiraishi2024,song2025,Kessler2024,Wang2026}. In parallel, a considerable effort has been devoted to other candidates, particularly NiAs structure (e.g.~MnTe and CrSb), which belong to a different symmetry class ($g$-wave) than the rutile ($d$-wave) materials.

A variety of experimental techniques have been used to detect altermagnetism, ranging from direct observation of spin splitting via spin-polarized angle-resolved photoemission spectroscopy, through measurements of magnon splitting reflecting the symmetry relations between magnetic sublattices, to charge-response probes such as the AHE and linear magneto-optical effects. While charge-response methods belong to the standard experimental toolbox and are closely connected to potential technological applications, they rely on the presence of spin–orbit coupling (SOC). The corresponding $\mathcal{T}$-odd response is governed by relativistic crystal symmetries and transforms as an axial vector (pseudovector), just like a net magnetization. Consequently, whenever a $\mathcal{T}$-odd charge response is symmetry allowed, a weak ferromagnetic moment is also permitted. Ferromagnetic and altermagnetic contributions therefore cannot be disentangled on symmetry grounds alone, as both enter through quantities with identical transformation properties.

X-ray magnetic circular dichroism (XMCD) is one such $\mathcal{T}$-odd response. For an x-ray wave vector $\mathbf{k}$, the XMCD signal is expressed as $2\,{\rm Im}\,\mathbf{h}(\omega)\!\cdot\!{\mathbf{k}}$, where the x-ray Hall vector $\mathbf{h}(\omega)$ is constructed from the antisymmetric part of the optical conductivity tensor~\cite{Hariki2024}. The Hall vector $\mathbf{h}(\omega)$ is itself a pseudovector and therefore transforms in the same way as the net magnetization. Accordingly, XMCD obeys the same symmetry constraints as the AHE. At x-ray frequencies, the dominant terms in the Hamiltonian generating the response involve the core orbitals. This leads to qualitatively different behavior from the AHE~\cite{Hariki2024,Hariki_mnf2,Kunes2025} and has motivated theoretical proposal that the altermagnetic and ferromagnetic contributions to the x-ray Hall vector $\mathbf{h}(\omega)$ can be distinguished by applying an external field~\cite{Hariki2025}. However, XMCD experiments have so far been limited to the model $g$-wave altermagnets $\alpha$-MnTe and Fe$_2$O$_3$, special cases where the ferromagnetic contribution is inherently negligible~\cite{Hariki2024,Amin2024,Yamamoto2025,Xie25}.

\begin{figure}[t]
	\begin{center}
		\includegraphics[width=0.99\columnwidth]{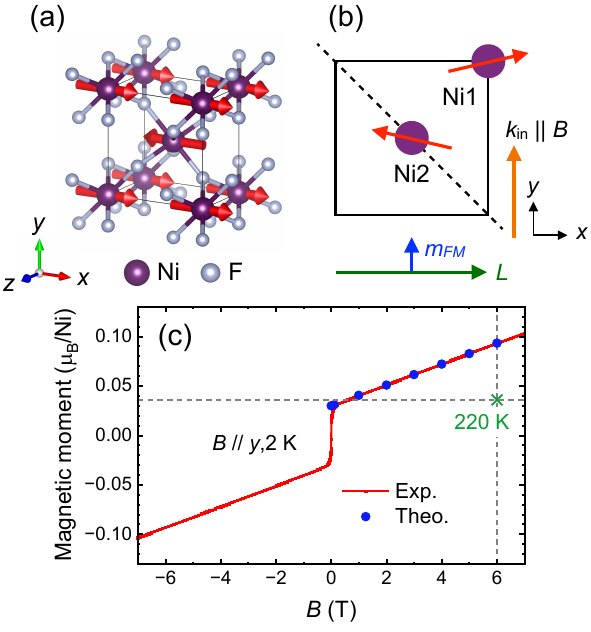}
	\end{center}
	\caption{(a) Crystal and magnetic structures of rutile NiF$_2$. Red arrows indicate the spin directions of the two antiferromagnetic sublattices on the Ni sites. Here, x-, y-, and z- represent the crystal's a, b, and c axis, respectively. (b) Schematic picture of the experimental geometry, where the incident x-ray wave vector $\boldsymbol{k}_{\rm in}$ is parallel to the applied field $\boldsymbol{B}$ and the sample $y$-axis. In this configuration, the net magnetic moment $\boldsymbol{m}_{\rm FM}$ aligns along $y$-axis and the N\'eel vector $\mathbf{L}$ points along $x$-axis. 
    The black dashed line represents the (110) surface of the sample.
    (c) Field dependence of the magnetization, measured by SQUID at 2~K with the magnetic field applied along the $y$-axis. The magnetization data point at 6~T and 220~K (green star) is shown which is extracted from the corresponding XMCD spectra. We also present the magnetization calculated within the present theory (blue dots).}
	\label{struct_mag}
\end{figure}

In this Letter, we present XMCD evidence of $d$-wave altermagnetism in rutile NiF$_2$ by experimentally demonstrating the unique capability of XMCD to disentangle the altermagnetic contribution from coexisting SOC- and field-induced ferromagnetism. NiF$_2$ is a wide-gap Mott insulator that exhibits  N\'eel order below $T_{\rm N}=73.2$~K~\cite{Stout1953} with $\mathbf{L}\parallel\langle 100\rangle$~\cite{Erickson1953,Matarrese1954} shown in Fig.~\ref{struct_mag}(a). Owing to the spin canting, a small in-plane net moment of about 0.03~$\mu_{\rm B}$~\cite{Matarrese1954,Moriya1960} is induced perpendicular to N\'eel vector $\mathbf{L}={\bf m}_1-{\bf m}_2$, defined as the difference between the sublattice magnetizations ${\bf m}_{1,2}$, as illustrated in Fig.~\ref{struct_mag}(b). Unlike $\alpha$-MnTe, where the altermagnetic contribution dominates in the absence of an external magnetic field due to  its extremely small net moment of $10^{-3}$--$10^{-4}\,\mu_{\rm B}$~\cite{Hariki2024, Kluczyk2024}, both the altermagnetic and ferromagnetic contributions are sizable in NiF$_2$. We show that the XMCD signal can, to very good approximation, be decomposed into a simple sum of altermagnetic and ferromagnetic contributions, $\Delta_{\rm ALT}(\omega)$ and $\Delta_{\rm FM}(\omega)$, which are extracted directly from experimental data acquired at two selected magnetic fields. We further propose that an alternative experimental determination of these functions can be achieved using XMCD measurements across the Néel temperature.

\begin{figure}[t]
	\begin{center}
		\includegraphics[width=0.99\columnwidth]{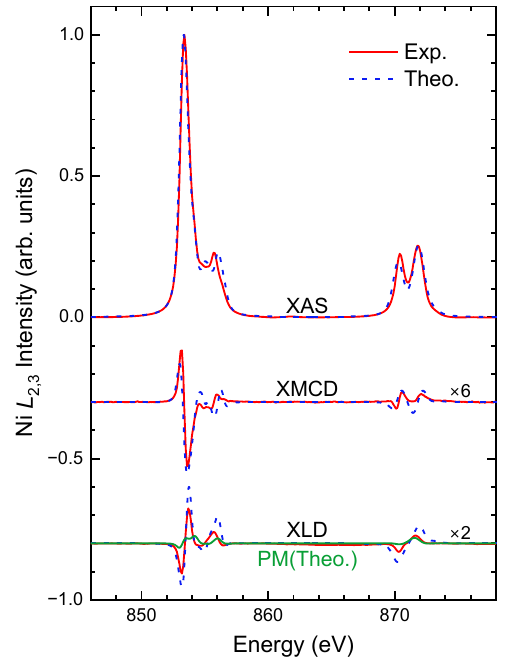}
	\end{center}
    \caption{Ni $L_{2,3}$-edge XAS (top), XMCD (middle), and XLD (bottom) spectra measured at 2~K under an applied field of 0.1~T. Red solid lines show the experimental data, and blue dashed lines show the calculated spectra. The XMCD and XLD spectra are scaled by constant factors indicated in the figure. The theoretical XLD spectrum in the paramagnetic (PM) phase is also shown (green line). The calculated spectral intensities at the $L_3$ ($L_2$) edge are broadened by a Lorentzian of 0.20~eV (0.25~eV) and a Gaussian of 0.20~eV (0.35~eV) (HWHM).}
	\label{xmcd1}
\end{figure}

\begin{figure*}[t]
	\begin{center}
		\includegraphics[width=1.99\columnwidth]{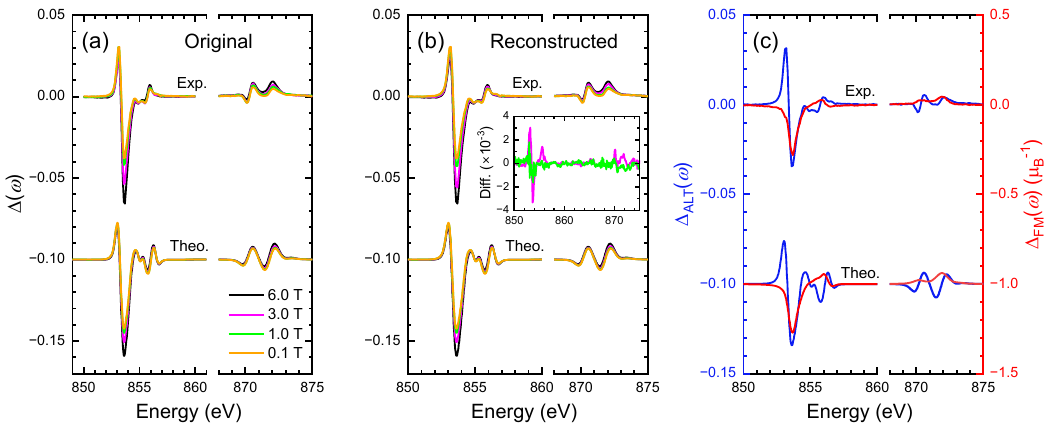}
	\end{center}
    \caption{(a) Ni $L_{2,3}$-edge XMCD spectra measured at different magnetic fields up to 6~T (top). The corresponding simulated spectra are shown at the bottom. The XMCD is defined as $\Delta(\omega)=(\mu_{+}(\omega)-\mu_{-}(\omega))/(\mu_{+}(\omega)+\mu_{-}(\omega))_{\rm max}$. (b) XMCD spectra reconstructed as a linear combination of the extracted (c) $\Delta_{\rm ALT}(\omega)$ (blue, left axis) and $\Delta_{\rm FM}(\omega)$ (red, right axis) according to Eq.~\eqref{eq:h}. The inset in (b) shows the difference between the experimental XMCD spectra and the reconstructed spectra.}
	\label{xmcd2}
\end{figure*}

The Ni $L_{2,3}$-edge x-ray absorption spectroscopy (XAS) measurements were performed on the I06 beamline at Diamond Light Source using total electron yield (TEY) detection.  The experimental geometry is shown in Fig~\ref{struct_mag}(b). The magnetic field was applied parallel to the incident beam and the sample $y$-axis. For the XMCD study, the XAS spectra were measured using circular polarised light with the photon helicity parallel ($\mu_+$) and antiparallel ($\mu_-$) to the applied field. For the x-ray linear dichroism (XLD) measurements, the XAS spectra were recorded by switching between vertical and horizontal linear polarization ($l_v$ and $l_h$) under an applied field of 0.1~T, corresponding to $\mathbf{E}\parallel x$ and $\mathbf{E}\parallel z$, respectively. All XAS, XMCD, and XLD measurements were performed at 2~K unless otherwise stated explicitly. Magnetization measurements were conducted using a Quantum Design MPMS SQUID-VSM magnetometer at 2~K.

The magnetization at 2~K (Fig.~\ref{struct_mag}(c)) exhibits a jump-wise increase
at low fields
which reflects the transition from four magnetic domains to a single domain with
the weak ferromagnetic moment aligned along the magnetic field~\cite{Matarrese1954,Moriya1960,Borovik1973}. Above 0.1~T a linear dependence without any hysteretic behaviour indicates
the system is in a single domain.
Magnetic susceptibility measurements shown in Fig.~\ref{mt} in SM
revealed the onset of magnetic ordering at about 73~K in agreement
with literature~\cite{Matarrese1954,Moriya1960,Borovik1973}.

The Ni $L_{2,3}$ XAS spectra at 2~K and 0.1~T are shown in Fig.~\ref{xmcd1}. The XAS spectra exhibit characteristic multiplet features of Ni$^{2+}$, similar to those of NiO~\cite{Alders1998}. A sizable XMCD (solid red) with a relative magnitude of approximately 5\% is observed. 
The XMCD exhibits a complex oscillatory line shape with sign inversions at both the $L_3$ and $L_2$ edges.
An oscillatory XMCD was observed and interpreted as a signature of altermagnetism 
in $\alpha$-MnTe~\cite{Hariki2024,Amin2024,Yamamoto2025}, however the line shape of NiF$_2$ is distinctly different. First, NiF$_2$ contains an additional ferromagnetic contribution, which will be addressed below. Second, the two systems have different $d$ fillings (Ni$^{2+}$ with $d^8$, Mn$^{2+}$ with $d^5$) and crystal structures, and therefore possess distinct multiplet structures that influence the $L_{2,3}$ XAS line shape~\cite{groot90}. Furthermore, the symmetry analysis of Refs.~\onlinecite{Hariki_mnf2,Kunes2025} revealed different roles of the core–valence multiplet (exchange) interaction in the XMCD of the two compounds.
Thus, a material-specific simulation is necessary to interpret the XMCD.

We simulate the XMCD using a Ni$^{2+}$ ionic model following Ref.~\onlinecite{Hariki2025}. The model incorporates crystal-field parameters derived from DFT calculations, as well as SOC and full multiplet interactions. To determine the effective fields acting on the Ni sublattices in the altermagnetic state with a canted moment, an antiferromagnetic Heisenberg Hamiltonian on the rutile lattice is solved within a static mean-field approximation.

Beyond Ref.~\onlinecite{Hariki2025}, we establish a quantitative altermagnetic model coexisting with weak ferromagnetism by determining three key parameters for this rutile $e_g$-orbital system: the non-cubic splitting between the $e_g$ orbitals ($\Delta_{e_g}$), the SOC constant $\xi_{3d}$, and the exchange parameter $J_{\rm ex}$. First, these parameters are constrained to reproduce the SQUID magnetization under applied fields, as shown in Fig.~\ref{struct_mag}(c). As detailed in the SM~\cite{sm}, fitting the magnetization data up to 6~T largely fixes $\xi_{3d}$ and $J_{\rm ex}$, leaving $\Delta_{e_g}$ as the only remaining parameter. We find that the XMCD is sensitive to $\Delta_{e_g}$, allowing us to determine a realistic value of $\Delta_{e_g} \approx -32$~meV. The resulting spectra (dashed blue) in the altermagnetic state with the experimental ferromagnetic moment show excellent agreement with the XAS and XMCD data.

We next disentangle the altermagnetic, $\Delta_{\rm ALT}(\omega)$, and ferromagnetic, $\Delta_{\rm FM}(\omega) m_{\rm FM}$, contributions to XMCD by verifying that the magnetic-field dependence of the XMCD spectra follow
\begin{equation}
\label{eq:h}
   \Delta(\omega;B)=\Delta_{\rm ALT}(\omega)+\Delta_{\rm FM}(\omega) m_{\rm FM}(B)
\end{equation}
where $m_{\rm FM}$ is the measured magnetization per atom in the magnetic field $B$, as shown in Fig.~\ref{struct_mag}(c). Note that the above formula is a special case of the general expression in Ref.~\cite{Hariki2025} for the present geometry with the magnetic field perpendicular to the easy axis. The linear decomposition is not guaranteed by the general symmetry arguments and therefore requires the experimental verification. Also, its applicability to NiF$_2$ has so far been supported only by numerical simulations and relies on the smallness of SOC in the Ni $3d$ shell 
as well as on additional, likely less significant, approximations.

To confirm the approximately linear relationship, we acquired XMCD spectra as a function of magnetic field up to 6~T shown in Figure~\ref{xmcd2}(a). Based on Eq.~\eqref{eq:h}, the $\Delta_{\rm ALT}(\omega)$ and $\Delta_{\rm FM}(\omega)$ are extracted using the data at 0.1~T and 6~T (Fig.~\ref{xmcd2}(c)). 
Calculations under the corresponding external fields are also performed for the theoretical model developed above, and the same procedure is applied to extract these fundamental spectra. As shown in Fig.~\ref{xmcd2}(a) and (b), an excellent agreement between experiment and theory is found for both $\Delta_{\rm ALT}(\omega)$ and $\Delta_{\rm FM}(\omega)$. The two components are then used to generate the reconstructed XMCD spectra at other field strengths, shown in Fig.~\ref{xmcd2}(b). The reconstructed XMCD spectra exhibit remarkable similarity to the corresponding experimental data with negligible deviations shown in the inset of Fig.~\ref{xmcd2}(b).


\begin{figure}[t]
	\begin{center}
		\includegraphics[width=0.9\columnwidth]{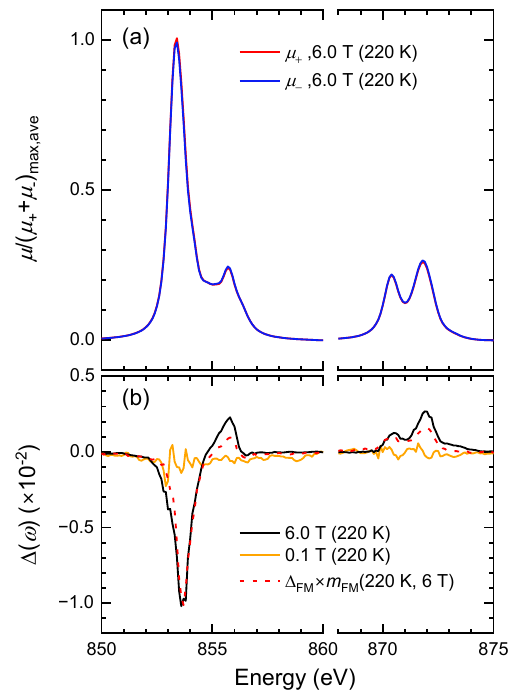}
	\end{center}
	\caption{(a) Ni $L_{2,3}$-edge $\mu_+$ (red) and $\mu_-$ (blue) spectra measured at 220~K under magnetic fields of 6.0~T. (b) XMCD spectra measured at 220~K (well above the N\'eel temperature) under magnetic fields of 6.0~T (black) and 0.1~T (orange). The red dashed line represents $\Delta_{\rm FM}(\omega)$ scaled by the magnetic moment $m_{\rm FM}(B=6~\rm {T},220~\rm {K})=0.036~\mu_{\rm B}$ extracted from the XMCD spectrum at 220~K under a 6~T magnetic field using the sum rule.} 
	\label{xmcd_220k}
\end{figure}


An alternative route can be taken to obtain $\Delta_{\rm FM}(\omega)$ and  $\Delta_{\rm ALT}(\omega)$. Figure~\ref{xmcd_220k} reports the Ni $L_{2,3}$ XMCD recorded at $T=220$~K, well above the N\'eel temperature of 73.2~K in external magnetic fields of 0.1~T and 6~T. The XMCD amplitude 
at 0.1~T is negligibly small (orange) demonstrating that both the altermagnetic and the ferromagnetic contribution $\Delta_{\rm ALT}(\omega)$ are zero above $T_{\rm N}$. In a large magnetic field of 6~T, a sizable XMCD signal (black) appears due to the induced magnetic moment.
A conjecture that the paramagnetic XMCD signal is given by $\Delta_{\rm FM}(\omega) m$, where $m$ denotes the field-induced magnetic moment, is quantitatively confirmed by the comparison shown in Fig.~\ref{xmcd_220k}. The measured XMCD at 6T (black curve) agrees closely with $\Delta_{\rm FM}(\omega) m$ (red dashed curve), where the latter is calculated using the paramagnetic moment $m = 0.036\mu_{\rm B}$ extracted from the XMCD spectrum via the sum rules (see SM~\cite{sm} for details of the analysis).
The very good match of $\Delta_{\rm FM}(\omega) m$ obtained below and above $T_{\rm N}$
is not {\it a priori} guaranteed since the
two arise from averaging over different sets of states. However, in the present case of a Mott insulator the atomic nature of the local state survives in the paramagnetic phase. Taken together with the frequency dependence of the Hall vector being approximately the same for all orientations of the local moment~\cite{Hariki2024}
the observation is not surprising.

In Fig.~\ref{xmcd1}, we present the XLD spectra (solid red) in addition to the XMCD one. A large XLD, in a good agreement with the theoretical simulations (dashed blue),
is observed in the magnetically ordered phase. Calculations in the absence of magnetic order, i.e., in the paramagnetic phase, yield only a weak XLD (solid green) signal. This indicates that the non-cubic crystal field plays a minor role for XLD. Indeed, the observed magnetic XLD is consistent with a magnetically ordered Ni$^{2+}$ ion in the cubic crystal field~\cite{Arenholz2007}. Further numerical results presented in Fig.~\ref{oh_d2h} of the SM~\cite{sm} confirm that the non-cubic crystal-field components including $\Delta_{e_g}$ has a negligible effect on XLD spectra.

The present results underscore the utility of XMCD for investigating altermagnets, particularly in comparison with other charge-response probes such as the AHE or the magneto-optical Kerr effect (MOKE), which probe valence and conduction states. The key advantage of XMCD, which additionally involves core states, lies in the distinct role of valence SOC, here associated with the Ni $3d$ shell.

In AHE or MOKE, valence SOC is an essential ingredient; without it, the dichroic effects vanish by symmetry. In contrast, for XMCD—where excitations from core $p$, $d$, or $f$ states are involved—the necessary relativistic symmetry is provided by the strong core SOC. The role of valence SOC is therefore comparatively minor and can be separated into two distinct effects:
(i) determining the orientation of the local moments, and
(ii) modifying the energy spectra and wave functions beyond simple spin-space rotations.

Numerical results for several $3d$ compounds~\cite{Hariki2024,Hariki_mnf2} indicate that contribution (ii) is nearly negligible in the XMCD spectra (see Fig.~\ref{delta_alt_wosoc3d} of the Supplemental Material for the present case). Effect (i), on the other hand, amounts to selecting the orientation of the Néel vector $\mathbf{L}$ from the non-relativistic manifold and possibly inducing a small ferromagnetic moment due to canting. In the local picture of NiF$_2$, this corresponds to selecting the appropriate local ground state within the $S=1$ subspace without substantially altering the spectra or wave functions.
The orientation of the magnetic moments is crucial for XMCD to be symmetry-allowed. Stating that the role of valence SOC is minor therefore means that essentially the same XMCD spectra are obtained when valence SOC is replaced by other mechanisms that fix the local moment orientation. In theory, this may be imposed as a mathematical constraint. Experimentally, the orientation of local moments can be tuned by applying a magnetic field, as in the present work, or by applying a mechanical strain.

Eq.~\eqref{eq:h} is a special instance of more general formula~\cite{Hariki2025}, which generalizes
the non-relativistic relationship between the XMCD signal and the orientation of the N\'eel vector~\cite{Kunes2025,Hariki_mnf2} in some $d$-wave altermagnets. Experimental confirmation of the accuracy of the Eq. (\ref{eq:h}) opens the possibility of exploiting the continuous variation of the N\'eel vector~\cite{Hariki_mnf2} in scanning XMCD experiments~\cite{Amin2024,Yamamoto2025}. Moreover, a number of altermagnets with rutile structure, e.g., all $3d$ dichalcogenides besides NiF$_2$, possess $c$-easy axis, which does not admit AHE, MOKE or XMCD. The reorientation of the N\'eel in order 
to allow these responses may be achieved by an external magnetic field~\cite{Feng2022}. The present
results shows that separation of the altermagnetic contribution to the XMCD signal is possible
and the electronic structure calculation substantiate its accurate description.

Last but not least, the comparison between the experimental and theoretical XMCD, XLD and XAS data reveals a different sensitivity to the microscopic parameters. The XLD in the magnetically ordered state is governed by cubic component of the crystal field 
with no sensitivity to the non-cubic splitting  $\Delta_{e_g}$ of $e_g$ states. This does not apply to XMCD which is sensitive to $\Delta_{e_g}$,
the parameter characterizing the local anisotropy in the rutile altermagnet with active $e_g$ orbitals. Our experimental results, together with a refined set of parameters, provide an accurate description of the altermagnetism in NiF$_2$, which is accompanied by a small intrinsic ferromagnetic component.

In conclusion, we reveal $d$-wave altermagnetism in rutile-structure NiF$_2$ using x-ray magnetic circular dichroism (XMCD). Although weak ferromagnetism induced by relativistic spin-orbit coupling is present and, by symmetry, contributes to the XMCD response, we demonstrate that the altermagnetic and ferromagnetic contributions naturally decouple in the XMCD response and can be accurately extracted through simple field- or temperature-dependent measurements, in excellent agreement with theoretical simulations. 
This result opens new opportunities for unambiguously characterizing altermagnets coexisting with a weak but non-negligible ferromagnetism, a situation frequently encountered in real materials.


\begin{acknowledgments}
We thank Diamond Light Source for providing beamtime under proposal ID MM40366-1 and the fruitful discussions with Sarnjeet Dhesi. Y.F.C. acknowledges funding from Diamond Light Source and the Clarendon Scholarship from the University of Oxford under joint doctoral studentship no. STU0477.
This work was also supported by JSPS KAKENHI Grant Numbers 25K00961, 25K07211, 23H03816, 23H03817, the 2025 Osaka Metropolitan University (OMU) Strategic Research Promotion Project (Young Researcher) (A.H.), the project Quantum materials for applications in sustainable technologies (QM4ST),
funded as project No. CZ.02.01.01/00/22 008/0004572 by Programme Johannes Amos Commenius, call Excel-
lent Research (J.K.) and by the Ministry of Education, Youth
and Sports of the Czech Republic through the e-INFRA
CZ (ID:90254) (A.H., J.K.). S.W.C. was supported by the DOE under Grant No. DOE: DE-FG02-07ER46382. C. W. was supported by the National Research Foundation of Korea(NRF) funded by the Ministry of Science and ICT(No. RS-2022-NR068223). We would like to acknowledge U.K. Engineering Physical Science Research Council (EPSRC grant no. EP/T028637/1) and Oxford-ShanghaiTech Collaboration project for their financial support. 

\end{acknowledgments}

\bibliography{nif2}

\clearpage
\onecolumngrid   

\setcounter{figure}{0}
\renewcommand{\thefigure}{S\arabic{figure}}

\section*{Supplementary Material of "X-ray magnetic circular dichroism evidence of \\ intrinsic $d$-wave altermagnetism in rutile-structure NiF$_2$"}

\subsection{Experimental Details}

\subsubsection{Crystal growth}
Commercially available NiF$_2$ powder (>99\% purity) was purified to remove moisture and oxygen by heating under a flow of high-purity argon gas (99.99\%) at 450 °C for 24 h. The purified powder was then loaded onto a graphite boat, and crystals were grown using a vapor growth method. Crystal growth was carried out in a tube furnace at 1200 $^o$C under a flowing argon atmosphere (100 cc min$^{-1}$) for two weeks~\cite{YUWU1976}. Appropriate safety measures were implemented to prevent vapour leakage and cross-contamination.

\subsubsection{Laue orientation of single crystal}

\begin{figure}[h]
	\begin{center}
		\includegraphics[width=0.6\columnwidth]{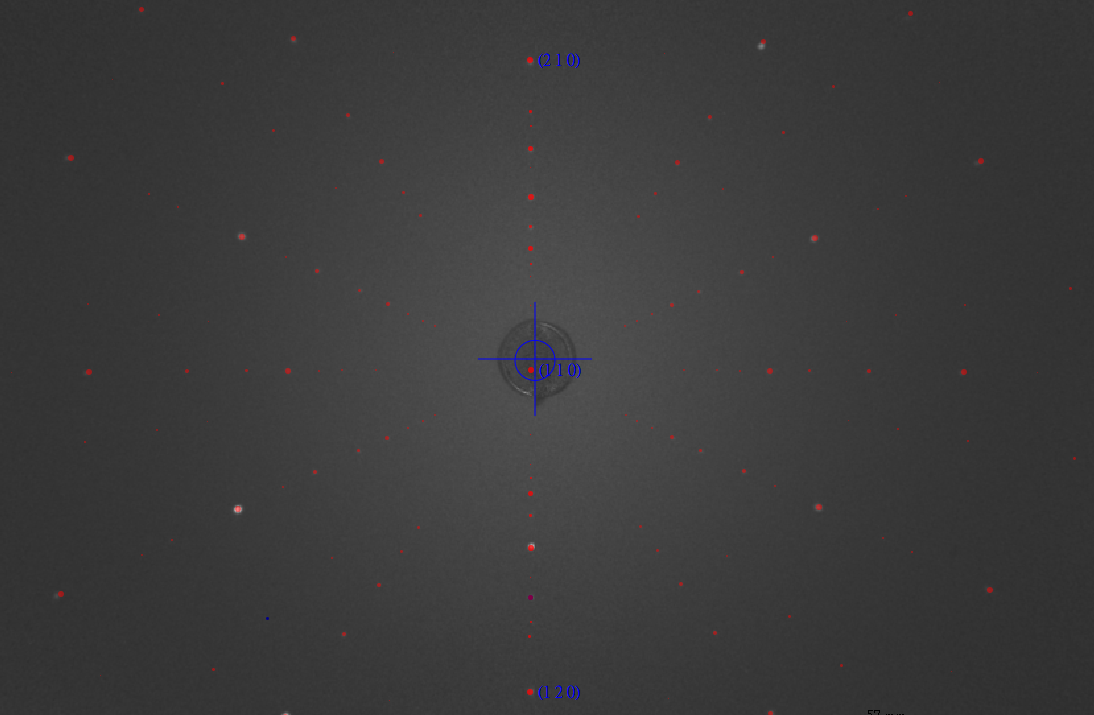}
	\end{center}
	\caption{X-ray Laue of NiF$_2$ sample with [110] surface normal.} 
	\label{laue}
\end{figure}

The crystallographic orientation of the single crystal was determined using back-reflection Laue x-ray diffraction at room temperature prior to the XMCD measurements. The Laue patterns were indexed using a simulated pattern based on the known crystal structure, confirming that the sample surface normal is aligned within ±1° of the [110] direction. The in-plane crystallographic axes were determined with an uncertainty of less than 1° (Fig.~\ref{laue}).
\\

\subsubsection{Magnetization measurements}

\begin{figure}[h]
	\begin{center}
		\includegraphics[width=0.5\columnwidth]{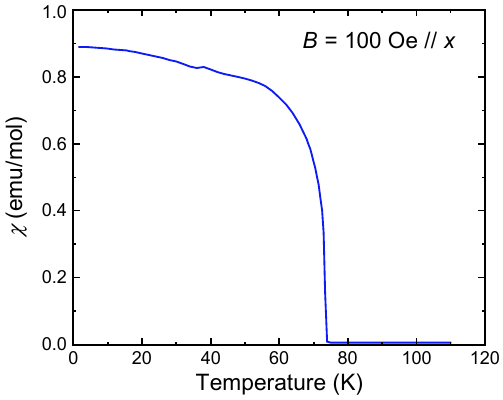}
	\end{center}
	\caption{Temperature dependence of the magnetic susceptibility $\chi(T)$ measured under field-cooled (FC) conditions in an applied magnetic field of $H$ = 100 Oe parallel to the $x$-axis ($H$~$\parallel$~$x$).}
	\label{mt}
\end{figure}

Magnetization measurements were conducted using a Quantum Design MPMS SQUID-VSM magnetometer. The temperature dependence of the magnetization was measured under field-cooled (FC) conditions with a magnetic field of 100 Oe applied parallel to the $x$-axis. As shown in Fig. \ref{mt}, the magnetization exhibits a rapid increase below 73~K, consistent with the onset of long-range magnetic ordering at $T_{\rm N}$~$\approx$ ~ 73~K. The ferromagnetic-like increase at low temperatures suggests the presence of a weak ferromagnetic component arising from spin canting.

\subsubsection{Ni L$_{2,3}$ edge XMCD}
The Ni $L_{2,3}$-edge XMCD measurements were performed on the I06 beamline at Diamond Light Source using total electron yield (TEY) detection, with an energy resolving power of $E/\Delta E > 10000$. The NiF$_2$ single crystal was mounted with the sample surface corresponding to the (110) plane.  The (110) sample surface was capped with a 1 nm Pt layer to prevent charging during TEY acquisition. The X-ray incidence angle was set to 45$^\circ$ with respect to the sample surface to maintain the incident beam direction and the applied magnetic field parallel to the sample $b$-axis (Fig. \ref{struct_mag} (b)). For the XMCD study the XAS was measured with photon spin parallel ($\sigma^+$) or antiparallel  ($\sigma^-$) with respect to the magnetic field. The degree of circular polarization delivered by the Apple II undulator sources was 99.8\% for the Ni $L$ edge. To improve signal-to-noise and suppress systematic effects from sample drift during measurements, each spectrum was collected as the average of multiple repeated scans, acquired in an alternating-helicity sequence (e.g., $+,-,-,+,\cdots$). The $\mu_{+}$ and $\mu_{-}$ spectra were constructed by combining opposite field and helicity configurations,
\begin{equation}
	\begin{aligned}
		\mu_{+}=\left[\mu(\sigma^+,+B)+\mu(\sigma^-,-B)\right]/2, \\
		\mu_{-}=\left[\mu(\sigma^{-},+B)+\mu(\sigma^{+},-B)\right]/2
	\end{aligned}
\end{equation}

For the XLD study the XAS was measured with linearly polarized light coming in with the electric field vector $\textbf{E}$ normal and parallel to the $xy$ plane. A weak magnetic field of 0.1~T was applied along the $x$-axis to ensure the system was in a single domain state.

\subsection{Experimental data analysis}

\subsubsection{Background treatment for XAS spectra}

\begin{figure}[t]
	\begin{center}
		\includegraphics[width=0.4\columnwidth]{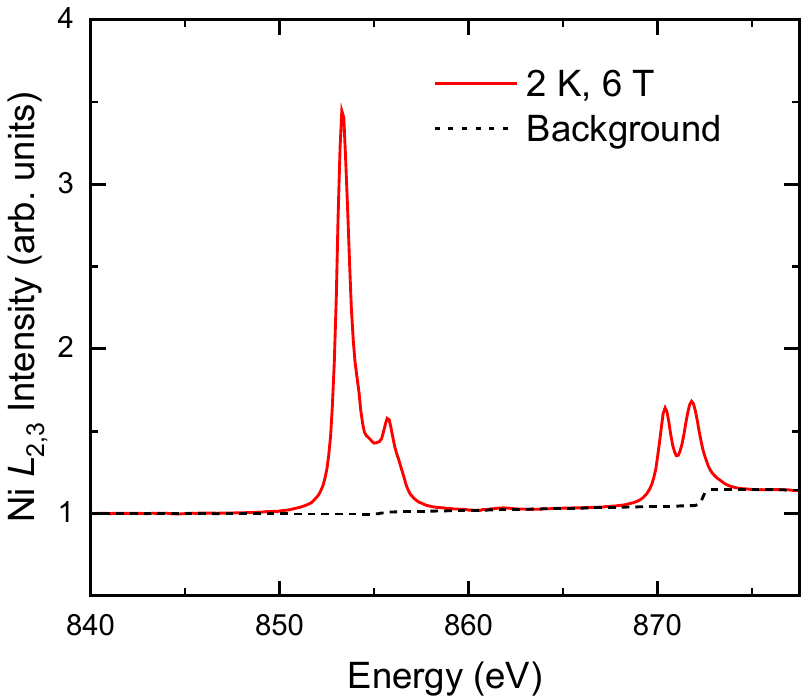}
	\end{center}
	\caption{Background correction for the XAS spectra. The red solid line shows a representative raw XAS spectrum ($\mu^{+}$) acquired at 2 K and 6 T. The black dashed line is the fitted background.}
	\label{bg}
\end{figure}

\begin{figure}[t]
	\begin{center}
		\includegraphics[width=0.99\columnwidth]{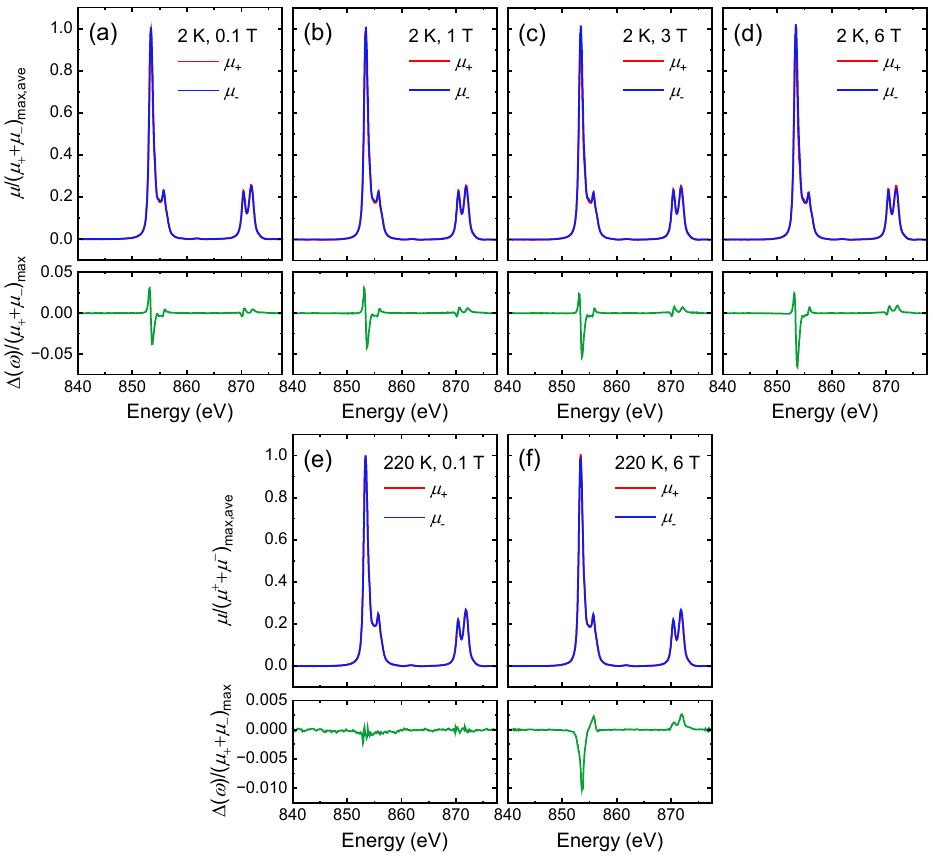}
	\end{center}
	\caption{Background subtracted and normalized $\mu_{+}$ and $\mu_{-}$ XAS spectra and the corresponding XMCD measured at 2 K under magnetic fields of 0.1, 1.0, 3.0, and 6.0 T [(a)–(d)], and at 220 K under 0.1 and 6.0 T [(e) and (f)].} 
	\label{XAS}
\end{figure}

The raw XAS spectra were corrected for background contributions. First, a linear baseline was fitted to the pre-edge region (840--845 eV) and subtracted. Then, the step-like continuum contributions across the $L_3$ and $L_2$ edges were removed by subtracting two arctangent step functions. For each edge, the step center energy was set to 2 eV above the corresponding white-line maximum, and the step width was fixed to 0.3 eV to account for lifetime broadening. Finally, a linear function was fitted to the high-energy post-edge region (876--878 eV), and this fitted line was subtracted. Fig.\ref{bg} shows an example raw XAS spectrum measured at 2 K and 6 T together with the fitted background. After background subtraction, the XAS spectra $\mu_{+}$ and $\mu_{-}$ were normalized to the maximum of $\mu_{+}+\mu_{-}$. Fig. \ref{XAS} present the background corrected and normalized $\mu_{+}$ and $\mu_{-}$ spectra at 2 K [(a)-(d)] and 220 K [(e)-(f)], respectively.

\subsubsection{Sum rule analysis}

The spin and orbital magnetic moments can be obtained by the sum-rules on the XMCD spectra. For the $L_{2,3}$ edge of transition metals, the sum rules have the form:
\begin{align}
	m_{\mathrm{orb}}
	&= -\frac{4 \displaystyle\int_{L_3+L_2}\!\left(\mu_{+}-\mu_{-}\right)\,d\omega}
	{3 \displaystyle\int_{L_3+L_2}\!\left(\mu_{+}+\mu_{-}\right)\,d\omega}
	\langle n_h\rangle,
	\\
	m_{\mathrm{s}} + 7\langle T_z\rangle
	&= -\frac{
		6 \displaystyle\int_{L_3}\!\left(\mu_{+}-\mu_{-}\right)\,d\omega
		-4 \displaystyle\int_{L_3+L_2}\!\left(\mu_{+}-\mu_{-}\right)\,d\omega}
	{\displaystyle\int_{L_3+L_2}\!\left(\mu_{+}+\mu_{-}\right)\,d\omega}
	\langle n_h\rangle.
\end{align}
Here, $n_h$ is the number of holes in the 3$d$ orbitals and $m_{\rm s}$ + 7$\langle T_z\rangle$ is the effective spin moment.

The sum-rule analysis was performed independently for each repeated measurement. The reported moments are the mean values over $N$ repetitions, $\overline{m}=\frac{1}{N}\sum_{i=1}^{N} m_i$, with the dispersion quantified by the sample standard deviation $s=\sqrt{\frac{1}{N-1}\sum_{i=1}^{N}(m_i-\overline{m})^2}$. The statistical uncertainty of the mean was estimated as the standard error $\mathrm{SE}=s/\sqrt{N}$ with a 95 \% confidence interval taken into account.

\begin{figure}[t]
	\begin{center}
		\includegraphics[width=0.99\columnwidth]{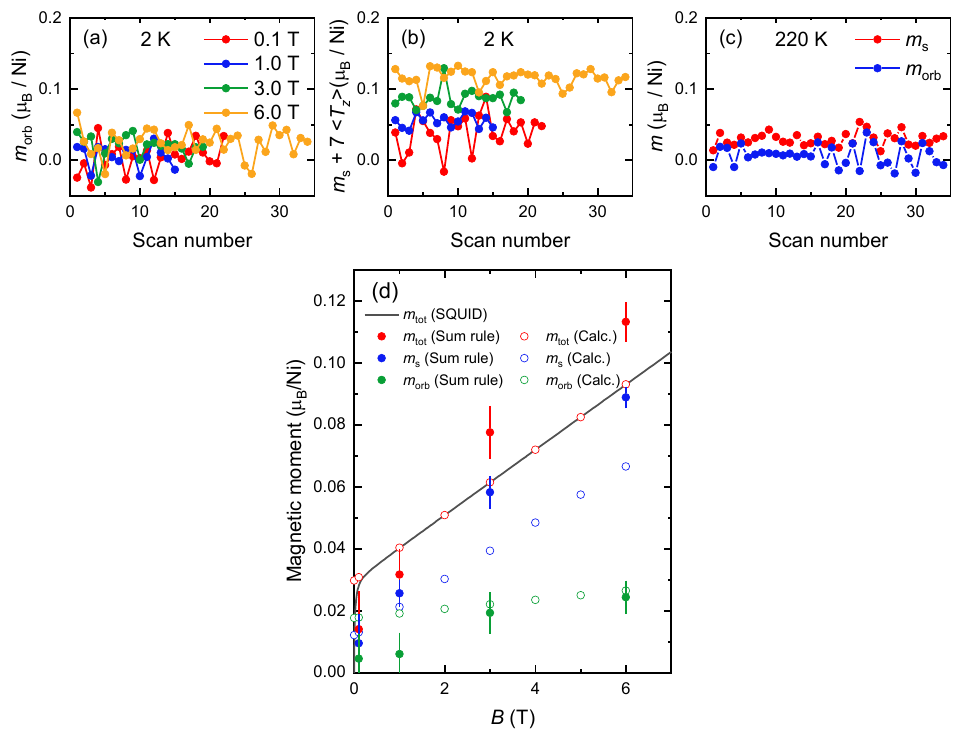}
	\end{center}
	\caption{(a) Orbital moments $m_{\rm orb}$ and (b) effective spin moments $m_{\rm s} + 7 \langle T_z \rangle$ obtained from XMCD sum-rule analysis for repeated measurements at 2 K. (c) Orbital moments $m_{\rm orb}$ and spin moments $m_{\rm s}$ obtained from XMCD sum-rule analysis for repeated measurements at 220 K. Each data point in (a)-(c) corresponds to the value extracted from an individual repeated measurement. (d) Magnetization curve of NiF$_2$ (black solid line), together with the calculated total magnetic moment $m_{\rm tot}$ and its spin and orbital components ($m_{\rm s}$ and $m_{\rm orb}$) shown by open circles. Experimental $m_{\rm tot}$, $m_{\rm s}$, and $m_{\rm orb}$ obtained from the XMCD sum rules are also shown by filled circles.}
	\label{sumrule}
\end{figure}

Fig.~\ref{sumrule} (a)-(b) shows the $m_{\rm s}$ and $m_{\rm orb}$ values extracted from each individual repeated measurement at 2 K for different magnetic fields. The corresponding mean values with error bars of $m_{\rm s}$ and $m_{\rm orb}$ are summarized in Fig.~\ref{sumrule} (d). An analogous analysis for 220K is presented in Fig. \ref{sumrule} (c). At 220~K, we obtain $m_{\rm s}=0.0298 \pm 0.0027~\mu_{\rm B}$/Ni and $m_{\rm orb}=0.0063 \pm 0.0042~\mu_{\rm B}$/Ni.

The calculated total magnetic moment $m_{\rm tot}$ and its spin and orbital components ($m_{\rm s}$ and $m_{\rm orb}$) are also reported in Fig.~\ref{sumrule} (d) as dashed lines. The values of the moments given by the application of the sum rules are fairly in agreement with the calculated ones. Small differences between experiment and calculated moments are expected due to the ionic model considered in the theory, which does not include the hybridisation of Ni $3d$ - F $2p$ orbitals. Hybridisation with the ligands is well known to induce a partial reduction of the orbital to spin moment ratio. Considering that the values of the parameters in the theory were optimised by $m_{\rm tot}$ (red dashed) reproducing the experimental magnetisation (solid black), the calculations may, hence, slightly underestimate $m_{\rm s}$ and overestimate $m_{\rm orb}$. 
In the application of the sum rules we considered $n_h = 2$ number of 3$d$ holes assuming a purely ionic Ni$^{2+}$ 3$d^8$ ground state. However, $n_h$ will be somewhat reduced in the real material because of the hybridisation, which might explain the values of $m_{\rm tot}$ given by the sum rules being larger of the magnetisation measured by the squid.  

\subsubsection{Spectral decomposition: $\Delta_{\rm ALT}$ and $\Delta_{\rm FM}$}

Based on the Eq.~(\ref{eq:h}), the altermagnetic component $\Delta_{\rm ALT}(\omega)$ and ferromagnetic component $\Delta_{\rm FM}(\omega)$ are extracted using XMCD spectra at two representative fields, 0.1~T and 6~T. Under the linear approximation, the $\Delta_{\rm FM}(\omega)$ corresponds to the derivative $\partial \Delta \mu/\partial m$. Using the two field data, the $\Delta_{\rm FM}(\omega)$ is obtained by:
\begin{equation}
	\Delta_{\rm FM}(\omega) = \frac{\partial \Delta \mu}{\partial m}~\simeq~\frac{\Delta \mu^{(6\rm T)}(\omega)-\Delta \mu^{(0.1\rm T)}(\omega)}{m^{(6\rm T)}-m^{(0.1\rm T)}} 
\end{equation}
Here, the magnetic moment $m(B)$ is extracted from the XMCD sum-rule analysis.

With $\Delta_{\rm FM}(\omega)$ determined, the $\Delta_{\rm ALT}(\omega)$ is obtained from the 6~T XMCD spectrum by subtracting the ferromagnetic contribution:
\begin{equation}
	\Delta_{\rm ALT}(\omega) = \Delta \mu^{(6\rm T)}(\omega) - m^{(6\rm T)}\Delta_{\rm FM}(\omega)
\end{equation}

Finally, the XMCD spectra can be reconstructed using the Eq.~(\ref{eq:h}) as shown in Fig. \ref{xmcd2}(b).

\clearpage

\subsection{Theoretical simulation}

\begin{figure*}[t]
	\begin{center}
		\includegraphics[width=0.95\columnwidth]{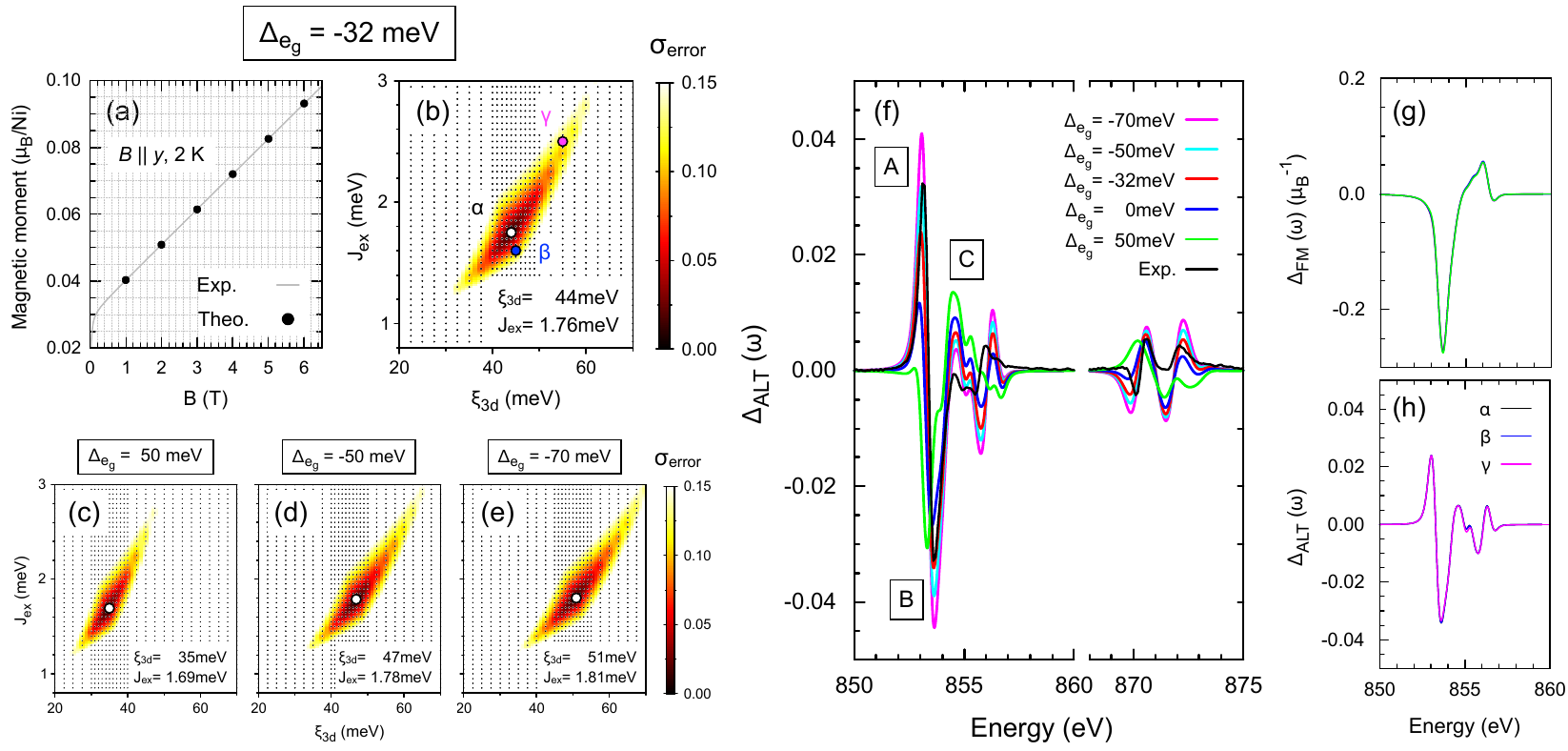}
	\end{center}
\caption{(a) Fit to the experimental magnetization data for $\Delta_{e_g}=-32$~meV. (b) Fitting error $\sigma_{\rm error}$ of the magnetization data as a function of $\xi_{3d}$ and $J_{\rm ex}$ for $\Delta_{e_g}=-32$~meV, and the corresponding results for (c) $\Delta_{e_g}=50$~meV, (d) $\Delta_{e_g}=-50$~meV, and (e) $\Delta_{e_g}=-70$~meV. The optimal values of $\xi_{3d}$ and $J_{\rm ex}$ are given in each panel. (f) Simulated $\Delta_{\rm ALT}(\omega)$ for selected $\Delta_{e_g}$ with optimized $\xi_{3d}$ and $J_{\rm ex}$ values from panels (b)--(e). The dotted points indicate the values at which the calculations are performed. Characteristic features $A$, $B$, and $C$ in the $L_3$ region are indicated. The fitting error $\sigma_{\rm error}$ is defined as the maximum relative deviation from the experimental value, $\Big|\dfrac{m^{\rm Theo.}(B)-m^{\rm Exp.}(B)}{m^{\rm Exp.}(B)}\Big|$, evaluated for the simulated field values $B$, where $m^{\rm Theo.}(B)$ and $m^{\rm Exp.}(B)$ denote the theoretical and experimental magnetization, respectively. (g) Corresponding $\Delta_{\rm FM}(\omega)$. (h) Simulated $\Delta_{\rm ALT}(\omega)$ for $\Delta_{e_g}=-32$~meV with optimal and non-optimal values of $\xi_{3d}$ and $J_{\rm ex}$ indicated in panel (a).}
	\label{sm_theo_param}
\end{figure*}  

Ni $L$-edge XMCD spectra are calculated using a Ni$^{2+}$ ionic model. A Weiss field $\mathbf{b}$ is introduced to simulate the altermagnetic ordered state coexisting with weak canted ferromagnetism. Computational details are provided in Ref.~\cite{Hariki2025}. In the present work, the model parameters are optimized by incorporating the experimental magnetization data under applied fields and the XMCD spectra. In Ref.~\cite{Hariki2025}, the crystal-field parameters were estimated from DFT calculations based on the experimental crystal structure. The spin–orbit coupling (SOC) constant $\xi_{3d}$ and the Slater integrals of the atomic Coulomb multiplet interaction are set to typical values for Ni$^{2+}$ systems. The nearest-neighboring spin-exchange coupling $J_{\rm ex}$ in the Heisenberg Hamiltonian $\mathcal{H}=-J_{\rm ex}\sum_{\langle i,j\rangle}{\mathbf{S}_i\cdot\mathbf{S}_j}$ is introduced following earlier studies on NiF$_2$~\cite{Moriya1960}. To determine the effective fields $\mathbf{b}$ acting on the Ni sublattices, the spin Hamiltonian is solved within a static mean-field approximation, where the local spin at each Ni site is obtained by diagonalizing the Ni ionic Hamiltonian. A uniform external field is included in the mean-field calculations when comparing with experimental data measured under applied magnetic fields. Once a converged mean-field solution for a N\'eel state coexisting with canted ferromagnetism is achieved, the XMCD spectra are computed by explicitly including the Ni $2p$ core orbitals and their full multiplet interaction with the Ni $3d$ orbitals~\cite{Hariki2020}.

Starting from the parameter set obtained in the previous work, we refine three parameters: (1) the splitting of the $e_g$ orbitals $\Delta_{e_g}$, (2) the SOC constant $\xi_{3d}$, and (3) the exchange parameter $J_{\rm ex}$. For Ni$^{2+}$ with a $d^8$ configuration, the splitting within the $t_{2g}$ manifold plays only a minor role in the quantities discussed here. As a first step, for a given parameter set $(\Delta_{e_g}, \xi_{3d}, J_{\rm ex})$, we calculate the evolution of the magnetization up to 6~T by performing mean-field calculations that incorporate the external magnetic field, as shown in Fig.~\ref{sm_theo_param}(a). We then search for an optimal parameter set that reproduces the experimental magnetization up to 6~T. For computational feasibility, the calculations are performed from 0.1~T to 6~T in steps of 1~T. Although three parameters are introduced, we find that for a given $\Delta_{e_g}$ the values of $(\xi_{3d}, J_{\rm ex})$ are nearly uniquely determined, as shown in Figs.~\ref{sm_theo_param}(b)-(e) for selected values of $\Delta_{e_g}$, effectively reducing the optimization to a single-parameter problem in terms of $\Delta_{e_g}$. The resulting values of $\xi_{3d}$ and $J_{\rm ex}$ that yield the best fit to the magnetization depend only weakly on $\Delta_{e_g}$, as shown in each panel.

The optimal value of $\Delta_{e_g}$ is then determined by reproducing the experimental XMCD profile at 0.1~T, where the signal is dominated by $\Delta_{\rm ALT}(\omega)$. As shown in Fig.~\ref{sm_theo_param}(f), $\Delta_{\rm ALT}(\omega)$ is sensitive to $\Delta_{e_g}$, allowing its optimal value to be determined within a relatively narrow range from $-30$~meV to $-50$~meV. In particular, $\Delta_{e_g}$ controls the amplitude of the distinct features marked $A$ and $B$, and to some extent feature $C$. Feature $C$ appears sharper in the theoretical spectrum, consistent with what is already observed in the XAS spectrum in Fig.~2 of the main text. We attribute this discrepancy to the present model, which does not include hybridization with ligand $p$ bands explicitly; it is well known that for Ni$^{2+}$ compounds such high-energy features broaden beyond the ionic approximation, for example within cluster or Anderson impurity models. The ferromagnetic contribution does not depend on $\Delta_{e_g}$, see Fig.~\ref{sm_theo_param}(g). Based on this analysis, we adopt $(\Delta_{e_g}, \xi_{3d}, J_{\rm ex})=(-32, 44, 1.76)$~meV in the main text. As mentioned above, $J_{\rm ex}$ and $\xi_{3d}$ do not play a decisive role at this stage of the optimization; nevertheless, for completeness, Fig.~\ref{sm_theo_param}(h) shows that $\Delta_{\rm ALT}(\omega)$ is largely insensitive to variations in $J_{\rm ex}$ and $\xi_{\rm SOC}$, where the simulations are performed with fixed $\Delta_{e_g}$ but with non-optimal values of $(\xi_{3d}, J_{\rm ex})$ indicated in Fig.~\ref{sm_theo_param}(b).

\begin{figure*}[t]
	\begin{center}
		\includegraphics[width=0.40\columnwidth]{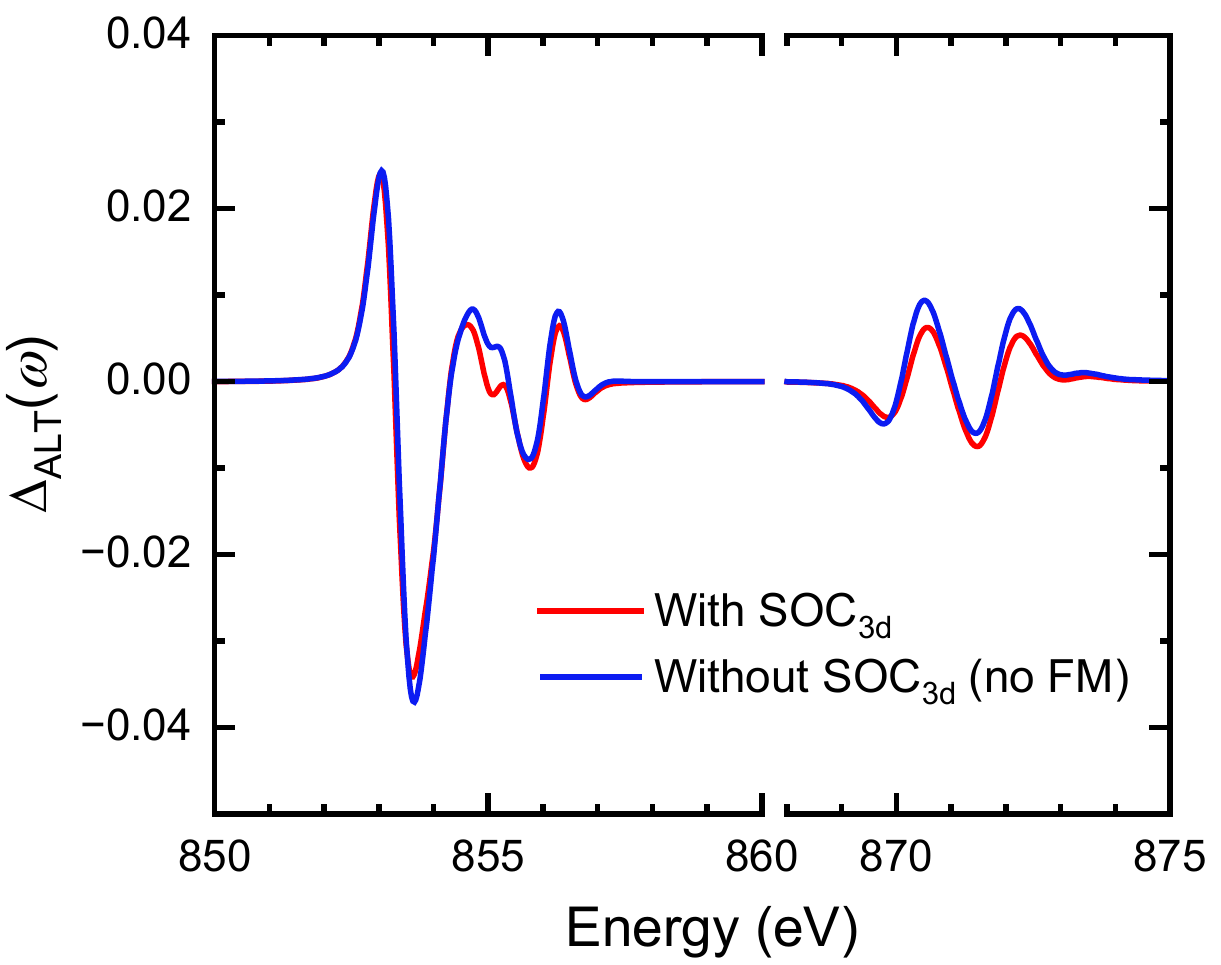}
	\end{center}
    \caption{Comparison of $\Delta_{\rm ALT}(\omega)$ calculated with (red) and without (blue) SOC on the Ni 3$d$ shell.}
	\label{delta_alt_wosoc3d}
\end{figure*}

Additionally, we calculated the XMCD using the optimized parameter set with SOC switched off in the Ni 3$d$ valence shell, while retaining SOC in the Ni 2$p$ core level, shown in Fig.~\ref{delta_alt_wosoc3d}. The SOC in the 2$p$ core is active only in the XAS final state and therefore does not influence the ground state. In this limit, the ferromagnetic moment in the ground state vanishes in the mean-field solution. The resulting XMCD spectrum is compared with the $\Delta_{\rm ALT}(\omega)$ derived in the main text. Apart from minor details in the spectral features, $\Delta_{\rm ALT}(\omega)$ is essentially insensitive to the presence or absence of SOC in the valence shell.

\begin{figure*}[t]
	\begin{center}
		\includegraphics[width=0.80\columnwidth]{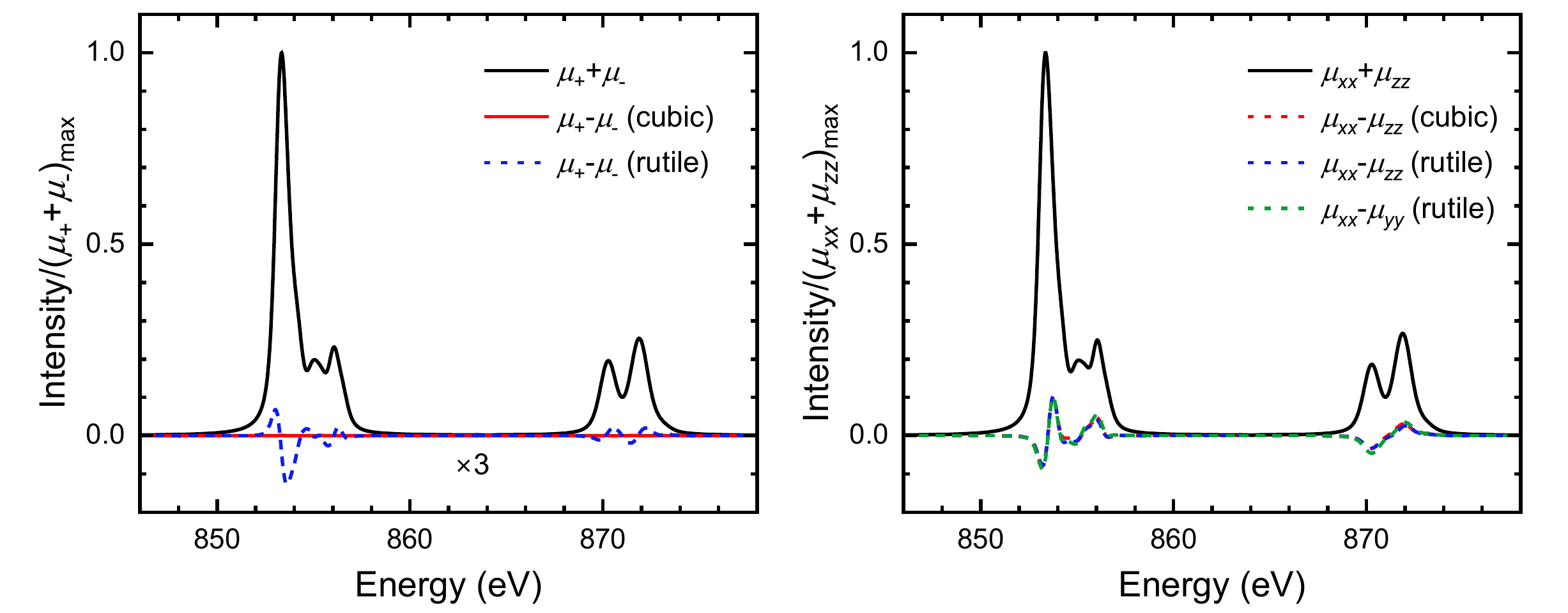}
	\end{center}
    \caption{Comparison of the simulated XMCD (left) and XLD (right) spectra. In both panels, the spectra shown in Fig.~\ref{xmcd1} for the rutile phase are reproduced (blue dashed lines). For comparison, the $\mu_{xx}-\mu_{yy}$ component is also shown (see discussion in the main text). The red lines correspond to results obtained in a cubic limit, where non-cubic contributions to the crystal-field terms are artificially eliminated in the Ni$^{2+}$ ionic Hamiltonian. The XMCD spectra are scaled by a factor of three.}
	\label{oh_d2h}
\end{figure*}

In Fig.~\ref{oh_d2h}, we compare the $\mu_{xx}-\mu_{zz}$ component shown in Fig.~\ref{xmcd1} with the $\mu_{xx}-\mu_{yy}$ component. In addition, we compute the XMCD and XLD spectra in a cubic limit, where non-cubic crystal-field components are set to zero both in the mean-field calculation and in the subsequent evaluation of the spectral intensities. Here, not only $\Delta_{e_g}$ but also all splittings within the $t_{2g}$ manifold are set to zero. We have verified that the splitting within the $t_{2g}$ manifold has a negligible effect on the XLD. 
It is clear that the $\Delta_{e_g}$ are crucial for inducing altermagnetic XMCD spectral from Fig~\ref{sm_theo_param} while the XLD is mainly due to the magnetic ordering rather than the non-cubic crystal-field components.

\end{document}